\documentclass[12pt]{iopart}

\usepackage{iopams}
\usepackage{mybbold}

\begin{document}

\eqnobysec

\noindent
ULM-TP/99-6 \\
HPL-BRIMS-1999-06 \\
September 1999\\

\jl{1}

\title{Semiclassical form factor for chaotic systems with spin 1/2}

\author{Jens Bolte{\rm \dag}%
\footnote[4]{E-mail address: {\tt bol@physik.uni-ulm.de}}  
and Stefan Keppeler{\rm \ddag \S}%
\footnote[5]{E-mail address: {\tt kep@physik.uni-ulm.de}}\footnote[6]
{Address after 1 October 1999: Abteilung Theoretische Physik, 
Universit\"at Ulm, Albert-Einstein-Allee 11, D-89069 Ulm, Germany}}

\address{\dag\ Abteilung Theoretische Physik, Universit\"at Ulm, 
Albert-Einstein-Allee 11, D-89069 Ulm, Germany}

\address{\ddag\ School of Mathematics, University of Bristol, 
University Walk, Bristol BS8 1TW, United Kingdom}

\address{\S \ BRIMS, Hewlett-Packard Laboratories, Filton Road, 
Stoke Gifford, Bristol BS34~8QZ, United Kingdom}

\begin{abstract}
We study the properties of the two-point spectral form factor for 
classically chaotic systems with spin 1/2 in the semiclassical limit,
with a suitable semiclassical trace formula as our principal tool. To
this end we introduce a regularized form factor and  discuss the limit 
in which the so-called diagonal approximation can be recovered. The 
incorporation of the spin contribution to the trace formula requires an
appropriate variant of the equidistribution principle of long periodic 
orbits as well as the notion of a skew product of the classical translational 
and spin dynamics. Provided this skew product is mixing, we show that 
generically the diagonal approximation of the form factor coincides with 
the respective predictions from random matrix theory.
\end{abstract}



\maketitle

\section{Introduction}
\label{1sec}
One of the major paradigms of quantum chaos is the conjecture of Bohigas, 
Giannoni and Schmit (BGS) \cite{BohGiaSch84} which states that the local 
statistics of energy spectra of (generic) individual quantum systems, 
whose classical analogues exhibit (strongly) chaotic behaviour, can be well 
described 
by that of ensembles of large random matrices. The symmetry properties of 
the relevant matrix ensembles have to be chosen according to the symmetries 
of the quantum system under consideration. In case the system is invariant 
under time reversal and has integer total angular momentum its local 
eigenvalue statistics are conjectured to be that of the Gaussian Orthogonal 
Ensemble (GOE). If time reversal invariance is broken one expects local
statistics according to the Gaussian Unitary Ensemble (GUE).
 
However, if the total angular momentum of the system is half-integer 
and the system is invariant under time reversal all eigenvalues show 
Kramers' degeneracy \cite{Kra30,Wig32} and their statistics have to be compared
with the Gaussian Symplectic Ensemble (GSE). In the GOE- and in the GUE-case 
there is plenty of numerical evidence available in favour of the 
BGS-conjecture, see, e.g., \cite{BohGiaSch84,Boh91,Gut90}, whereas 
only few examples have been studied in the GSE-case as, e.g., in 
\cite{SchDieKusHaaBer88,CauGra89}. For quantum 
systems whose classical limit is integrable, i.e., which shows regular 
behaviour, one expects the local eigenvalue statistics to follow the laws 
of a Poisson process \cite{BerTab77}. For the analytical treatment 
semiclassical methods, in particular semiclassical trace formulae, have become 
the most important tools since Berry and Tabor \cite{BerTab77} investigated 
the behaviour of the spectral form factor for classically integrable systems 
by means of an appropriate trace formula. By making use of the Gutzwiller 
trace formula \cite{Gut71,Gut90,DuiGui75,Mei92,PauUri95}, Berry provided a 
semiclassical theory for the spectral form factor \cite{Ber85} of classically 
chaotic systems without spin. Based on the so-called diagonal approximation 
he could explain the semiclassical asymptotics of the form factor for small 
values of its argument, thus recovering the GOE- and GUE-behaviour, 
respectively. 

In this paper our aim is to show that Berry's semiclassical treatment of
the two-point form factor can be carried over to quantum systems with
spin 1/2, whose classical translational dynamics are chaotic.
We base our analysis on the semiclassical trace formula for the Dirac 
equation that we developed recently \cite{BolKep98,BolKep99}. In this 
trace formula the presence of spin is reflected in a modification of
the amplitudes with which the periodic orbits of the translational 
dynamics contribute. This modification arises from a spin dynamics that 
involves a `classical' spin precessing along the periodic orbits. The
central part of this paper therefore consists of calculating the effect 
of this spin contribution to the semiclassical form factor. For our analysis 
we use a two-point form factor whose definition differs slightly from the
one that is more commonly used in spectral statistics as, e.g., in 
\cite{Ber85}. We rather prefer the point of view adopted in 
\cite{RudSar96,Mar98}. Both definitions, however, are equivalent in the 
limit where infinitely many eigenvalues are taken into account. We also 
stress that both the form factor and the associated correlation function 
are distributions and hence have to be evaluated on suitable test 
functions. This approach makes a spectral average obsolete and enables 
one to state the BGS-conjecture, specialized to the form factor, in a 
precise manner. Moreover, the lacking self-averaging property discussed 
in \cite{Pra97} poses no difficulty in this context.

The paper is organized as follows. In section \ref{2sec} we introduce our 
definition of the spectral form factor for a finite part of the spectrum 
and discuss in which sense one can expect to recover the form factors given 
by random matrix theory. Section \ref{3sec} is devoted to the definition of 
a regularized form factor which we evaluate semiclassically using trace 
formulae for the Dirac as well as for the Pauli equation 
\cite{BolKep98,BolKep99}. We also invoke the diagonal approximation and 
briefly discuss its range of validity. In section \ref{4sec} a suitable 
version of the equidistribution principle of long periodic orbits
is used in order to obtain the semiclassical asymptotics of the diagonal 
form factor. In this context we employ the notion of a skew product of 
the translational and the spin dynamics, the ergodic properties of which 
determine the semiclassical asymptotics. Our principal results are then 
summarized in section 5. Namely, depending on the presence or absence of
quantum mechanical time reversal invariance, and provided the dynamics of 
the skew product is mixing, we can recover a GSE- or GUE-behaviour of the 
diagonal form factor, respectively. The relation between classical and 
quantum mechanical time reversal as well as the equidistribution of long 
periodic orbits are discussed in two appendices.
  
\section{The form factor for quantum systems with half-integer spin}
\label{2sec}

The two-point correlations of a discrete quantum spectrum are conveniently
measured by either the two-point correlation function $R_2$ or by the
two-point form factor $K_2$, which is related to $R_2$ through a
Fourier transform. Before defining these quantities one usually
{\it unfolds} the spectrum, i.e., the eigenvalues $E_k$ are rescaled
to $x_k$ such that the unfolded eigenvalues have a mean separation of
one. This means that the spectral density $d(x)$ of the unfolded spectrum 
allows for a separation
\begin{equation}
\label{unfolddens}
d(x) := \sum_k \delta (x-x_k ) = 1 + d_{fl}(x) \ ,
\end{equation}
such that 
\begin{equation}
\label{dfldef}
\frac{1}{2\Delta x}\int_{x-\Delta x}^{x+\Delta x} d_{fl}(y)\ \rmd y =
\frac{1}{2\Delta x}\,\#\{ k;\ x-\Delta x\leq x_k\leq x+\Delta x\} - 1
\end{equation}
vanishes as $x\to\infty$, $\Delta x\to\infty$, $\Delta x/x\to 0$. For
a finite part of the spectrum, containing $N$ unfolded eigenvalues enumerated
as $x_1,\dots,x_N$, one defines the two-level correlation function by 
\begin{equation}
\label{Rsub2N}
R_2 (s;N) := \frac{1}{N} \sum_{k,l\leq N} \delta (s-(x_k -x_l )) - 1 \ .
\end{equation}
Accordingly, the two-level form factor is defined as
\begin{equation}
\label{Ksub2N}
K_2 (\tau;N) := \int_{\mathbb R} R_2 (s;N)\,\rme^{-2\pi\rmi\tau s}\ \rmd s 
= \frac{1}{N} \sum_{k,l\leq N} \rme^{-2\pi\rmi\tau ( x_k -x_l )} - 
\delta (\tau)\ .
\end{equation}
Since both quantities are distributions, which is most clearly seen in
the case of the correlation function (\ref{Rsub2N}), one should evaluate
these on test functions $\phi\in C^\infty_0 (\mathbb R)$,
\begin{equation}
\label{Rsub2Ntest}
\eqalign{
\int_{\mathbb R}K_2 (\tau;N)\,\phi(\tau)\ \rmd\tau 
  &= \frac{1}{N}\sum_{k,l\leq N}\int_{\mathbb R}\phi(\tau)\,
     \rme^{-2\pi\rmi\tau ( x_k -x_l )} \rmd\tau - \phi(0) \\
  &= \frac{1}{N} \sum_{k,l\leq N}\hat\phi(2\pi( x_l -x_k)) - \phi(0) \\
  &= \int_{\mathbb R}R_2 (s;N)\,\hat\phi(2\pi s)\ \rmd s \ .}
\end{equation}
We remark that since the form factor is obviously 
even in $\tau$, it suffices to consider only even test functions $\phi$.
The convention for the Fourier transform that was used, and that will be
used in all of what follows, is
\begin{equation}
\label{FT}
\hat f(k) = \int_{\mathbb R}f(x)\,\rme^{\rmi xk}\ \rmd x \qquad \mbox{and} 
\qquad f(x) = \frac{1}{2\pi}\int_{\mathbb R}\hat f(k)\,\rme^{-\rmi xk}
\ \rmd k \ .
\end{equation}
After smearing with a test function all expressions occurring in 
(\ref{Rsub2Ntest}) are obviously finite. In this form a semiclassical
analysis of either the form factor or the correlation function can be
carried out. If $\phi$ is chosen non-negative, the left-hand
side of (\ref{Rsub2Ntest}) can also be viewed as the mean value of 
$K_2(\tau;N)$
when $\tau$ is drawn randomly with probability density $\phi$. Since we
always understand the form factor in the above sense, the absence of a
self-averaging property discussed in \cite{Pra97} is not essential for
our further considerations. 

For a given quantum Hamiltonian $\hat H$ the unfolding of its discrete
spectrum shall proceed in the following manner. We consider a spectral
interval
\begin{equation}
\label{IofhE}
I = I(E,\hbar) := [ E-\hbar\omega,E+\hbar\omega ] \ , \quad \omega > 0 \ ,
\end{equation}
that has no overlap with a possible essential spectrum of $\hat H$. Then
the condition $E_k\in I$ is equivalent to
\begin{equation}
\label{EsubkinI}
-\omega \leq \frac{E_k -E}{\hbar} \leq \omega \ .
\end{equation}
The number of eigenvalues contained in $I$,
\begin{equation}
\label{NsubI}
N_I := \# \{ k;\ E_k \in I \}\ , 
\end{equation} 
can be estimated semiclassically as 
\begin{equation}
\label{NIsemicl}
N_I \sim 2\hbar\omega \bar d(E) = \frac{\omega}{\pi}\,T_H(E)\ ,\quad
\hbar\to 0\ ,
\end{equation}
where $\bar d(E)$ denotes an appropriate mean spectral density and $T_H(E) 
:= 2\pi\hbar\bar d(E)$ is the {\it Heisenberg-time}. A convenient definition 
of $\bar d(E)$ can be derived from the semiclassical trace formula for
$\hat H$ in that it shall denote the contribution coming from the singularity
of $\Tr\exp[-\frac{\rmi}{\hbar}\hat H t]$ at $t=0$ to all polynomial orders 
in $\hbar$ see, e.g., \cite{DuiGui75,Mei92,PauUri95,BolKep99,Bol99}. The 
spectra that we are going to consider below are such that in 
the semiclassical limit $\hbar\to 0$ the Heisenberg-time tends to infinity, 
$T_H\to\infty$. For example, given $E$ in the gap of the essential spectrum 
of a Dirac-Hamiltonian $\hat H_D$, i.e., in typical cases $-mc^2 < E < mc^2$, 
the mean spectral density reads \cite{BolKep98,BolKep99}
\begin{equation}
\label{dbar}
\bar d(E) = 2\,\frac{{\rm vol}\,\Omega^+_E + {\rm vol}\,\Omega^-_E}
{(2\pi\hbar)^3}\,[1+\Or (\hbar) ]\ , 
\end{equation}
where $\Omega^\pm_E$ denote the hypersurfaces of energy $E$ in phase space
corresponding to the classical Hamiltonians 
\begin{equation}
\label{classHam}
H^\pm (\bi{p},\bi{x}) = e\varphi (\bi{x}) \pm \sqrt{\left(c\bi{p}-
e\bi{A}(\bi{x})\right)^2 + m^2 c^4}
\end{equation}
for relativistic particles of positive and negative kinetic energy,
respectively, in the static external electromagnetic fields generated by
the potentials $\varphi$ and $\bi{A}$. Therefore, in the 
semiclassical limit the number of eigenvalues in the interval $I$ increases, 
although its length $|I|=2\hbar\omega$ shrinks to zero. A completely 
analogous argument applies to Pauli-Hamiltonians \cite{BolKep99}.

We now define the unfolded spectrum through
\begin{equation}
\label{defoldef}
x_k := E_k \bar d(E) \qquad \mbox{and} \qquad x := E \bar d(E) \ .
\end{equation}
The condition $E_k\in I(E,\hbar)$ is hence equivalent to $x_k\in
[x-\Delta x,x+\Delta x]$, where $\Delta x :=\frac{\omega}{2\pi}T_H$. With 
this choice indeed $x\to\infty$, $\Delta x\to\infty$, such that $\Delta x/x 
= \hbar\omega/E\to 0$ in the semiclassical limit. In this context the 
quantity (\ref{dfldef}) reads
\begin{equation}
\label{unfoldcond}
\frac{1}{2\Delta x}\,\#\{k;\ x-\Delta x\leq x_k\leq x+\Delta x\} - 1 = 
\frac{\pi}{\omega T_H}\,N_I - 1 \ ,
\end{equation}
such that (\ref{NIsemicl}) ensures its vanishing in the semiclassical limit.

From now on we will choose the numbering of the eigenvalues $E_k$ and $x_k$, 
respectively, in such a way that the eigenvalues in $I$ are given by
\begin{equation}
\label{eigenno}
E_1 \leq E_2 \leq \dots \leq E_{N_I} \ .
\end{equation}
Changing the value of $\hbar$ therefore alters the numbering of the 
eigenvalues. As a consequence, the condition $E_k\in I$ is equivalent to 
$k\leq N_I$. At this place we recall that the semiclassical limit 
$\hbar\to 0$, or $T_H\to\infty$, implies $N_I\to\infty$. For the form 
factor (\ref{Ksub2N}) we now obtain 
\begin{equation}
\label{Ksub2NsubI}
\eqalign{
K_2 (\tau;N_I) &= \frac{1}{N_I}\sum_{E_k,E_l\in I}\rme^{-2\pi\rmi\tau
                  \bar d(E)(E_k -E_l)} - \delta (\tau) \\
               &= \left| \frac{1}{\sqrt{N_I}} \sum_k \chi_{[-\omega,\omega]}
                  \left(\frac{E_k -E}{\hbar}\right)\,\rme^{-\frac{\rmi}{\hbar}
                  \tau T_H E_k}\right|^2 - \delta (\tau) \ ,}
\end{equation}
where the $k$-sum extends over all eigenvalues of $\hat{H}$ and 
$\chi_{[-\omega,\omega]}$ denotes the characteristic function of 
the interval $[-\omega,\omega]$ that occurs due to the condition 
(\ref{EsubkinI}).

In the preceding discussion we tacitly assumed that the discrete spectrum 
of the quantum Hamiltonian carries no systematic degeneracies. In order to 
achieve such a situation one has to remove all symmetries. As opposed to 
geometric or internal symmetries that have to be realized by unitary 
representations of the respective symmetry groups, the time reversal 
operation must be implemented by an anti-unitary operator $\hat T$, see 
\cite{Wig32} and \ref{appTR}. For single particles of spin $s$ the square of 
$\hat T$ depends 
on $s$ being integer or half-integer in that $\hat T^2 =(-1)^{2s}$. In case 
the quantum system is time reversal invariant, i.e., $[\hat H,\hat T]=0$, 
and has half-integer spin this leads to {\it Kramers' degeneracy}
\cite{Kra30}: Since $\hat T^2 =-1$ implies that every vector $\psi\neq 0$ 
in the Hilbert space is orthogonal to $\hat T\psi$, all eigenvalues of 
$\hat H$ are (at least) two-fold degenerate, see \cite{Haa91,Meh91}
for details. 

Following the usual practice, we will remove Kramers' degeneracy
by replacing each degenerate pair $E_{2k} =E_{2k+1}$ of eigenvalues by one 
of its representatives. Thus the mean spectral density is lowered by a 
factor of two. In analogy to (\ref{defoldef}) the unfolding $E_k\mapsto
\tilde x_k$ of the so modified spectrum can therefore be achieved through 
the choice $\tilde x_k := x_k/2$. The modified form factor then reads
\begin{equation}
\label{ffmod}
\eqalign{
\widetilde K_2 (\tau;N) 
  &= \frac{2}{N} \sum_{k,l\leq N \atop k,l\ odd}\rme^{-2\pi\rmi\tau 
     (\tilde x_k -\tilde x_l)} - \delta (\tau) \\
  &= \frac{1}{2N} \sum_{k,l\leq N}\rme^{-2\pi\rmi\frac{\tau}{2}
     (x_k -x_l)} -\frac{1}{2} \delta \left( \frac{\tau}{2} \right) \\
  &= \frac{1}{2}\,K_2 \left( \frac{\tau}{2};N \right) \ .}
\end{equation}
In this setting the conjecture of Bohigas, Giannoni and Schmit 
\cite{BohGiaSch84} states that for individual (generic) classically chaotic 
quantum systems of particles with half-integer total spin and no  
unitary symmetries one should obtain
\begin{equation}
\label{BGSwithT}
\lim_{N\to\infty} \int_{\mathbb R} \widetilde K_2 (\tau;N)\,\phi(\tau)\ 
\rmd\tau \stackrel{!}{=} \int_{\mathbb R} K_2^{GSE}(\tau)\,\phi(\tau)\ 
\rmd\tau 
\end{equation}
for all test functions $\phi\in C_0^\infty (\mathbb R)$. Here $K_2^{GSE}$
denotes the two-point form factor of the Gaussian symplectic ensemble
of random matrix theory,
\begin{equation}
\label{ffGSE}
K_2^{GSE}(\tau) = \cases{ \case{1}{2}|\tau|-\case{1}{4}|\tau|
\log|1-|\tau|| &for $|\tau| \leq 2$\ , \\
1 &for $|\tau| \geq 2$\ ,}
\end{equation}
see, e.g., \cite{Meh91}. When time reversal invariance is lacking, the
respective conjecture reads
\begin{equation}
\label{BGSnoT}
\lim_{N\to\infty} \int_{\mathbb R} K_2 (\tau;N)\,\phi(\tau)\ \rmd\tau 
\stackrel{!}{=} \int_{\mathbb R} K_2^{GUE}(\tau)\,\phi(\tau)\ \rmd\tau \ , 
\end{equation}
where now the form factor of the Gaussian unitary ensemble \cite{Meh91} 
should occur,
\begin{equation}
\label{ffGUE} 
K_2^{GUE}(\tau) = \cases{ |\tau| &for $|\tau| \leq 1$\ , \\
1 &for $|\tau| \geq 1$\ . } 
\end{equation}
In our subsequent semiclassical investigations we will in both cases, 
i.e., with and without time reversal invariance, consider the form factor
$K_2 (\tau;N_I)$ as it is given in (\ref{Ksub2NsubI}). When dealing with 
the case of time reversal invariance we appeal to the relation (\ref{ffmod}). 

\section{The semiclassical form factor}
\label{3sec}

Since the work of Berry and Tabor \cite{BerTab77} on the distribution
of eigenvalues for classically integrable systems, semiclassical trace 
formulae have found numerous and fruitful applications in the
analysis of spectral statistics. A prominent example is Berry's analysis 
of the spectral rigidity \cite{Ber85}, which relies in an essential way
on a semiclassical evaluation of the two-point form factor based on the 
Gutzwiller trace formula. In this work it already became apparent that 
present semiclassical methods at most allow to study the form factor in the 
restricted range $|\tau|< 1$, see also \cite{Bol99} for a review. Only 
recently, improved techniques have been developed \cite{BogKea96} that might 
allow to extend the semiclassical analysis of spectral statistics. In this 
paper, however, we follow the more traditional path in that in the end we 
consider the so-called diagonal approximation for the form factor.   

The two-point form factor as given in (\ref{Ksub2NsubI}) requires to 
establish a trace formula for the sum
\begin{equation}
\label{traceform}
\sum_k \chi_{[-\omega,\omega]}\left(\frac{E_k -E}{\hbar}\right)\,
\rme^{-\frac{\rmi}{\hbar}\tau T_H E_k} \ .
\end{equation}
However, the general structure of (convergent) semiclassical trace 
formulae, see, e.g., \cite{DuiGui75,Mei92,PauUri95,BolKep99,Bol99},
necessitates the use of a smooth test function $\rho\in C^\infty (\mathbb R)$
with Fourier transform $\hat\rho\in C_0^\infty (\mathbb R)$. One therefore
has to replace the sharp cut-off, provided by the characteristic function
in (\ref{traceform}), by a smoothened substitute. For this reason we now 
introduce the {\it regularized form factor}
\begin{equation}
\label{regformf}
K^{\chi,\eta}_2 (\tau;T_H) := \left| \sqrt{\frac{\pi}{\omega T_H}}\sum_k 
\chi(E_k)\,\eta\left(\frac{E_k -E}{\hbar}\right)\,\rme^{-\frac{\rmi}{\hbar}
\tau T_H E_k}\right|^2 - \delta (\tau) \ , 
\end{equation}
where $\eta\in C^\infty (\mathbb R)$ is a test function with Fourier
transform $\hat\eta\in C_0^\infty (\mathbb R)$, but that is otherwise
arbitrary at the moment. Later we will introduce a further normalization
condition. In the following we will consider both relativistic and
non-relativistic particles with spin 1/2. In the relativistic case, when
one is dealing with a Dirac-Hamiltonian $\hat H_D$, the function 
$\chi\in C_0^\infty (\mathbb R)$, which is not to be confused with the 
characteristic function $\chi_{[-\omega,\omega]}$,
is necessary to truncate the essential
spectrum of $\hat H_D$. In typical situations $\chi$ should therefore be 
supported in the interval $(-mc^2,mc^2)$, where the eigenvalues $E_k$ of 
$\hat H_D$ are located. When these do not accumulate at some point, one 
could also leave out the truncation $\chi$ from (\ref{regformf}). 

We are now in a position to use the test function
\begin{equation}
\label{tracetestfct}
\rho(\varepsilon) := \eta(\varepsilon)\,\rme^{-2\pi\rmi\bar d(E)\tau 
(\hbar\varepsilon +E)}
\end{equation}
in the semiclassical trace formula for the Dirac equation that was
developed in \cite{BolKep98,BolKep99},
\begin{equation}
\label{DiracTF}
\eqalign{
\sum_k \chi(E_k)\,\rho\left(\frac{E_k -E}{\hbar}\right)
  &= \chi(E)\,\frac{T_H (E)}{2\pi}\,\hat\rho (0) \, [1+\Or(\hbar)]   \\
  &\quad +\chi(E)\,\sum_{\gamma}\sum_{k\neq 0} \frac{T_\gamma}{2\pi}\,
     \hat\rho (kT_\gamma)\,A_{\gamma,k} \ .}
\end{equation}
The outer sum on the right-hand side extends over all primitive periodic 
orbits $\gamma$ of energy $E$, with periods $T_\gamma$, 
of the two classical flows 
generated by the Hamiltonians (\ref{classHam}). The inner sum then is over 
all $k$-fold repetitions of primitive orbits, formally including negative 
ones. The weight attached to each pair $(\gamma,k)$ reads
\begin{equation}
\label{amplit}
A_{\gamma,k} := \frac{\tr d_\gamma^k}
{|\det(\mathbb M_\gamma^{k} -\mathmybb{1})|^{\frac{1}{2}}} 
\, \rme^{\frac{\rmi}{\hbar}
kS_\gamma (E) - \rmi\frac{\pi}{2}k\mu_\gamma}\, [1+\Or(\hbar)] \ .
\end{equation} 
Here $d_\gamma \in$ SU(2) denotes the semiclassical time evolution operator
for the spin degrees of freedom along the primitive periodic orbit $\gamma$
of the translational dynamics. Furthermore, 
$S_\gamma (E)$ denotes the action of $\gamma$ and $\mu_\gamma$ is its Maslov 
index. The (monodromy) matrix $\mathbb M_\gamma$ is the linearized Poincar\'e 
map transversal to $\gamma$. In the form given in (\ref{DiracTF}) the trace 
formula is valid for all cases where the classical flows have only isolated 
and non-degenerate periodic orbits. An analogous trace formula, with 
appropriate simplifications, is also available for Pauli-Hamiltonians, see 
\cite{BolKep99}. 

Upon choosing the test function (\ref{tracetestfct}) in the trace formula 
(\ref{DiracTF}), its left-hand side reads
\begin{equation}
\label{lhsTF}
\sum_k \chi(E_k)\,\rho \left( \frac{E_k -E}{\hbar} \right) = \sum_k \chi
(E_k)\,\eta\left( \frac{E_k -E}{\hbar} \right)\,\rme^{-\frac{\rmi}{\hbar}
\tau T_H E_k} \ , 
\end{equation}
and is hence the appropriate starting point for a semiclassical analysis
of the form factor, compare (\ref{regformf}). As a first ingredient on
the right-hand side of (\ref{DiracTF}) one requires the Fourier transform
of the test function (\ref{tracetestfct}), which is given by
\begin{equation}
\label{Fourierrho}
\hat\rho (t) = \rme^{-\frac{\rmi}{\hbar}E\tau T_H}\,\hat\eta (t-\tau T_H)\ .
\end{equation}
For convenience we now choose $\eta$ to be even and real-valued, which 
implies that $\hat\eta$ also shares these properties. Furthermore, the 
truncation $\chi$ of the essential spectrum shall be such that $\chi(E)=1$. 
Thus, the trace formula yields the following semiclassical representation 
of the regularized form factor,
\begin{eqnarray}
\label{TFsquare}
\fl  
K^{\chi,\eta}_2 (\tau;T_H)
  =& -\delta (\tau)+\frac{T_H}{4\pi\omega} \left[\hat\eta(\tau T_H)
     \right]^2 \, [1+\Or(\hbar)]  \nonumber \\
  &+ \sum_\gamma\sum_{k\neq 0}\frac{T_\gamma}{4\pi\omega}\,\hat\eta
     (\tau T_H)\,\hat\eta(kT_\gamma -\tau T_H)\,A_{\gamma,k} \, 
     [1+\Or(\hbar)]  \\
  &+ \frac{1}{T_H}\sum_{\gamma,\gamma'}\sum_{k,k'\neq 0}
     \frac{T_\gamma T_{\gamma'}}{4\pi\omega}\,\hat\eta(kT_\gamma -\tau 
     T_H)\,\hat\eta(\tau T_H -k'T_{\gamma'})\,A_{\gamma,k}\,A_{\gamma',-k'}
     \ . \nonumber
\end{eqnarray}
In a next step we are going to test the semiclassical form factor with
some $\phi\in C^\infty_0 (\mathbb R)$, compare (\ref{Rsub2Ntest}). To this 
end one needs the integral
\begin{equation}
\label{Fdef}
F(T,T') := \int_{\mathbb R}\phi(\tau)\,\hat\eta(T-\tau T_H)\,
\hat\eta(\tau T_H- T')\ \rmd\tau \ ,
\end{equation}
whose leading term in the semiclassical limit $T_H\to\infty$ can be 
calculated by introducing the Fourier representations for $\phi$ and  
$\hat\eta$. A straight-forward calculation then yields
\begin{equation}
\label{testeval}
\fl
F(T,T') = \frac{1}{T_H}\int_{\mathbb R}\int_{\mathbb R}\int_{\mathbb R}
 \hat\phi(t)\,\eta(\varepsilon)\,\eta(\varepsilon')\,\rme^{\rmi
  (\varepsilon T -\varepsilon' T')} 
 \, \delta\left(\frac{t}{T_H}+\varepsilon -\varepsilon'\right)
  \ \rmd\varepsilon'\,\rmd\varepsilon\,\rmd t   \ .
\end{equation}
Changing variables from $\varepsilon,\varepsilon'$ to $u:=\varepsilon' 
-\varepsilon$ and $v:= (\varepsilon' +\varepsilon)/2$, and employing the 
expansions
\begin{equation}
\label{etaexp}
\eta\left(v\pm\frac{t}{2T_H}\right) = \eta(v) + 
\Or\left(\frac{t}{T_H}\right) \ ,
\end{equation}
finally shows that
\begin{equation}
\label{testeval1}
\eqalign{
F(T,T') 
 &= \frac{1}{T_H}\int_{\mathbb R}\int_{\mathbb R}\hat\phi(t)\,
    \rme^{-\rmi t(\frac{T+T'}{2T_H})}\,\eta(v)^2\,\rme^{\rmi v(T-T')}
    \ \rmd v\,\rmd t + \Or\left(\frac{1}{T_H^2}\right) \\
 &= \frac{1}{T_H}\,\phi\left(\frac{T+T'}{2T_H}\right)\,
    \hat\eta *\hat\eta (T-T') + \Or\left(\frac{1}{T_H^2}\right) \ , }
\end{equation}
where the convolution
\begin{equation}
\label{convo}
\hat\eta *\hat\eta (t) := \int_{\mathbb R}\hat\eta (t-t')\,\hat\eta(t')
\ \rmd t' = 2\pi\int_{\mathbb R}\eta (v)^2\,\rme^{\rmi tv}
\ \rmd v 
\end{equation}
enters. We therefore conclude that
\begin{equation}
\label{Regformtest}
\eqalign{ 
\fl
\int_{\mathbb R}K_2^{\chi,\eta}(\tau;T_H)\,\phi(\tau)\ \rmd\tau
  &= \phi(0) \left[-1 + \frac{1}{4\pi\omega}\,\hat\eta *\hat\eta(0)\right] 
     + \Or(\hbar) \\
  &\quad +\frac{1}{T_H}\sum_\gamma\sum_{k\neq 0}\frac{T_\gamma}{4\pi\omega}
     \,\hat\eta *\hat\eta (kT_\gamma)\,A_{\gamma,k}\,\phi \left(
     \frac{kT_\gamma}{2T_H}\right) \, [1+\Or(\hbar)]  \\
  &\quad +\frac{1}{T_H^2}\sum_{\gamma,\gamma'}\sum_{k,k'\neq 0}
     \frac{T_\gamma T_{\gamma'}}{4\pi\omega}\,\hat\eta *\hat\eta (kT_\gamma 
     -k'T_{\gamma'})  \\  
  &\hspace{2.5cm}\times\phi \left(\frac{kT_\gamma +k'T_{\gamma'}}{2T_H}
     \right)\,A_{\gamma,k}\,A_{\gamma',-k'} \ .  
}
\end{equation}
At this point we introduce the normalization of $\eta$ announced
previously. Guided by the simple observation 
\begin{equation}
\label{charnorm}
\int_{\mathbb R}[\chi_{[-\omega,\omega]}(\varepsilon)]^2 \ \rmd\varepsilon 
= 2\omega \ ,
\end{equation}
we require the same normalization for the smooth substitute $\eta$ of
the sharp cut-off $\chi_{[-\omega,\omega]}$,
\begin{equation}
\label{chinorm}
\frac{1}{2\pi}\,\hat\eta *\hat\eta (0) =  \int_{\mathbb R}\eta
(\varepsilon)^2\ \rmd\varepsilon \stackrel{!}{=} 2\omega \ .
\end{equation}
As a consequence, the leading semiclassical order of the first line on the 
right-hand side of (\ref{Regformtest}) vanishes. Furthermore, since the 
Fourier transform $\hat\eta$ of the test function $\eta$ is required to 
be compactly supported, the second line is a finite sum, multiplied by 
$1/T_H$. A similar argument applies to the third and fourth line, apart 
from the {\it diagonal contribution} with $kT_\gamma = k'T_{\gamma'}$, 
where $\hat\eta *\hat\eta(0)$ occurs and thus no such cut-off is present. 

Due to the above reasoning it is tempting to assume that in the 
semiclassical limit $T_H\to\infty$ the right-hand side of (\ref{Regformtest}) 
is completely fixed by the contribution of the {\it diagonal form factor}
\begin{equation}
\label{diagformf}
K_2^{diag}(\tau;T_H) := \frac{1}{T_H^2}\sum_{\gamma}\sum_{k\neq 0}
g_{\gamma,k}\,T_\gamma^2\,|A_{\gamma,k}|^2\,\delta\left(\tau-\frac{k
T_\gamma}{T_H}\right)\ .
\end{equation}
Here we assumed that $k'T_{\gamma'}=kT_\gamma$ implies $A_{\gamma',k'}=
A_{\gamma,k}$, see \ref{appTR} for the spin contribution,
and then $g_{\gamma,k}$ denotes the number of pairs 
$(\gamma,k)$ such that $kT_\gamma$ has a given value. It is, however, 
well known that the above assumption is not justified. The reason for 
this lies in the subtleties of the limits involved. In order to arrive 
at the left-hand side of (\ref{BGSwithT}), or of (\ref{BGSnoT}), one 
first has to remove the smoothening of the characteristic function 
$\chi_{[-\omega,\omega]}$ in that a sequence of functions 
$\eta\in C^\infty(\mathbb R)$ approaching
$\chi_{[-\omega,\omega]}$ has to be considered. Then, in the limit, 
$\hat\eta *\hat\eta$ is no longer compactly supported. Indeed, according
to (\ref{convo}) one obtains $\hat\eta *\hat\eta(t)=4\pi\,\sin(\omega t)/t$.
Still, the periodic-orbit sums in (\ref{Regformtest}) are truncated by
the test function $\phi$. However, in the semiclassical limit this cut-off 
is being removed. Moreover, for long periodic orbits the differences
$kT_\gamma -k'T_{\gamma'}$ can become arbitrarily small so that 
$\hat\eta *\hat\eta (kT_\gamma -k'T_{\gamma'})$ only provides a modest
truncation of near-diagonal contributions. The only example where it could 
be rigorously shown \cite{RudSar96} that the diagonal form factor itself 
produces the correct limit, if the test functions $\phi$ are restricted to 
those that are supported in the interval $[-1,1]$, is that of the 
correlations of the non-trivial zeros of principal $L$-functions, including 
the case of the Riemann zeta function, see also \cite{Mon73}. Rudnick and 
Sarnak \cite{RudSar96} even proved an analogous result for general $n$-point 
correlations. 

As announced previously, we now invoke the diagonal approximation, i.e.,
we leave aside the contribution of 
$K_2^{\chi,\eta}(\tau;T_H)-K_2^{diag}(\tau;T_H)$ 
to (\ref{Regformtest}). This procedure, which goes back to
Berry \cite{Ber85}, is generally supposed to reveal the correct behaviour
of the form factor for small $|\tau|$. The reason for this being that if
the test function $\phi$ is supported in a small interval, the contribution
of long periodic orbits to (\ref{Regformtest}) is truncated. Furthermore, 
due to the normalization (\ref{chinorm}) the diagonal form factor is 
independent of the smoothening $\eta$. This convenient fact exempts one 
from the need to discuss the removal of this smoothening. In order to test 
now the range of small $|\tau|$ one should restrict the class of test 
functions $\phi$ to those supported in intervals $[-\tau',\tau']$, where 
$\tau'>0$ is `small enough'. Recalling that $K_2^{diag}$ and $\phi$ are even 
in $\tau$ an integration by parts yields
\begin{equation}
\label{diagformtest}
\eqalign{
\fl
\int_{\mathbb R}K_2^{diag}(\tau;T_H)\,\phi(\tau)\ \rmd\tau 
&= 2 \int_0^{\infty} K_2^{diag}(\tau;T_H)\,\phi(\tau)\ \rmd\tau
\\
 &= -\int_{0}^{\tau'}\phi'(\tau)\,\frac{2}{T_H^2}\sum_{\gamma}
\sum_{k \geq 1 \atop kT_\gamma\leq \tau T_H}g_{\gamma,k}\,T_\gamma^2\,
|A_{\gamma,k}|^2 \ \rmd\tau\ . }
\end{equation}
What is required now is the asymptotic behaviour of the periodic-orbit
sum in (\ref{diagformtest}) as $T_H\to\infty$. Since the contributions of 
repetitions of primitive periodic orbits are asymptotically suppressed due to 
their stronger instabilities, compare also (\ref{DetMasy}) below, in the 
following we only take the $k=1$--term of the sum over the repetitions into 
account. As a consequence we therefore have to study the asymptotics of 
the periodic-orbit sum
\begin{equation}
\label{posumest}
\sum_{\gamma, \, T_\gamma\leq \tau T_H}\frac{g_{\gamma,1}
\,T_\gamma^2\,(\tr d_\gamma)^2}{|\det(\mathbb M_\gamma -\mathmybb{1})|}
\end{equation}
in the double limit $T_H\to\infty$, $\tau\to 0$, such that 
$\tau T_H\to\infty$. In order to simplify this task we now make two
assumptions, which should be verified in all cases that could be considered 
as `generic' in any reasonable sense: 
\begin{enumerate}
\item The periods $T_\gamma$ of primitive periodic orbits shall be such
that any finite subset of them is linearly independent over ${\mathbb Q}$.
This implies that the multiplicities $g_{\gamma,k}$ are independent of
$k$, i.e., $g_{\gamma,k}=g_\gamma$.
\item The subset of primitive periodic orbits $\gamma$ with $g_\gamma\neq
\bar g$ is of density zero in the set of all primitive periodic orbits,
\begin{equation}
\label{gbardense}
\lim_{T\to\infty}\frac{\#\{\gamma;\ g_\gamma\neq\bar g,\ T_\gamma\leq T\}}
{\#\{\gamma;\ T_\gamma\leq T\}} \stackrel{!}{=} 0 \ ,
\end{equation}
where $\bar g=2$ in case the classical dynamics are time reversal invariant,
and $\bar g=1$ when time reversal symmetry is absent. In this context time 
reversal invariance does not only mean that an orbit $\gamma$ is geometrically 
identical to its time reversed partner, but also that both orbits 
yield the same contribution of the spin degrees of freedom to the trace 
formula, which then implies that $A_{\gamma,k}$ is invariant under time 
reversal. In \ref{appTR} we show that this condition is a consequence of 
quantum mechanical time reversal invariance.
\end{enumerate}
Under these assumptions the factors $g_{\gamma,k}$ can be replaced by
$\bar g$ and can then be pulled out of the sum (\ref{posumest}). 

\section{Classical periodic-orbit sums and the contribution of spin}
\label{4sec}

The aim of this section is to obtain the leading semiclassical behaviour
of the periodic-orbit sum (\ref{posumest}). Apart from the appearance
of the Heisenberg-time only quantities related to the classical flow
enter this expression. It therefore seems appropriate to invoke
results about the distribution of periodic orbits in phase space. In
order to retain a certain convenient generality, we will not specify
the classical translational dynamics further, except for the following 
assumptions:
\begin{enumerate}
\item The classical flow $\Phi_H^t :\,\Omega_E\to\Omega_E$
on the compact hypersurface $\Omega_E$ of energy $E$ in the $2d$-dimensional 
phase space is generated by some Hamiltonian function $H(\bi{p},\bi{x})$ 
and hence preserves the (normalized) Liouville measure
\begin{equation}
\label{Lioum}
\rmd\mu_E (\bi{p},\bi{x}) := \frac{1}{{\rm vol}\,\Omega_E}\,\delta
(H(\bi{p},\bi{x})-E)\ \rmd^d p\,\rmd^d x 
\end{equation}
on $\Omega_E$.
\item $\Phi_H^t$ is {\it ergodic} with respect to Liouville measure.
\item $\Phi_H^t$ is {\it hyperbolic} on all of $\Omega_E$.
\end{enumerate}
In the case of the semiclassical form factor for a Dirac-Hamiltonian these
requirements shall apply to both classical flows, i.e., to those generated
by the two classical Hamiltonians $H^\pm$ given in
(\ref{classHam}). Moreover, we now assume that for a given energy $E$ there 
will only be either a contribution coming from the dynamics generated by 
$H^+$ or from the dynamics generated by $H^-$, but never from both at the 
same time. This is not a strong restriction since it only excludes situations 
in which Klein's paradox \cite{Kle29} can appear.

The hyperbolicity of the classical flows implies that in particular all
periodic orbits are either hyperbolic or loxodromic. This means that all
monodromy matrices $\mathbb M_\gamma$ have eigenvalues with moduli strictly
different from one. Since the eigenvalues occur in pairs of mutually inverse 
numbers, we denote them as $\rme^{\pm (u_{\gamma,j}+\rmi v_{\gamma,j})}$, 
$u_{\gamma,j}>0$, $v_{\gamma,j}\in [0,2\pi)$, $j=1,\dots,d-1$. Thus
\begin{equation}
\label{detMon}
\eqalign{
|\det(\mathbb M_\gamma -\mathmybb{1})| 
  &= \prod_{j=1}^{d-1} \left|\left(\rme^{u_{\gamma,j}+\rmi v_{\gamma,j}}
     -1\right)\,\left(\rme^{-u_{\gamma,j}-\rmi v_{\gamma,j}}-1\right)
     \right|  \\
  &= \exp\left( \sum_{j=1}^{d-1}u_{\gamma,j} \right)\,\prod_{j=1}^{d-1}
     \left| 1-\rme^{-u_{\gamma,j}-\rmi v_{\gamma,j}} \right|^2 \ . }
\end{equation}
The stability exponents $u_{\gamma,j}$ are related to the Lyapunov
exponents $\lambda_{\gamma,j}$ of $\gamma$ through $u_{\gamma,j} = 
\lambda_{\gamma,j}T_\gamma$ so that $u_{\gamma,j}\to\infty$ as 
$T_\gamma\to\infty$. Hence, in this limit one obtains
\begin{equation}
\label{DetMasy}
\frac{1}{|\det(\mathbb M_\gamma - \mathmybb{1})|} \sim p_\gamma :=
\exp\left( -\sum_{j=1}^{d-1}u_{\gamma,j} \right) \ .
\end{equation}
Since the semiclassical limit of the periodic-orbit sum (\ref{posumest})
is dominated by the contribution of  long periodic orbits, (\ref{DetMasy})
allows to analyze (\ref{posumest}) in terms of periodic-orbits sums
that are familiar from equidistribution theorems of periodic orbits, see,
e.g., \cite{ParPol90}.

For the kind of Hamiltonian flows characterized above mean values of 
observables on $\Omega_E$ with respect to Liouville measure can be
calculated with the help of appropriate periodic-orbit sums. We postpone
a detailed discussion of this matter to \ref{appsumrules}, from which we 
here only quote that for any continuous observable $a$ one obtains the 
representation
\begin{equation}
\label{ameanbypo}
{\bar a}^E := \int_{\Omega_E} a(\bi{p},\bi{x})\  \rmd\mu_E 
(\bi{p},\bi{x}) = \lim_{T\to\infty}\frac{1}{T}\sum_{\gamma,\,T_\gamma\leq T}
T_\gamma\,{\bar a}^\gamma\,p_\gamma\ ,
\end{equation}
where ${\bar a}^\gamma$ denotes an average of $a$ along the periodic orbit 
$\gamma$,
\begin{equation}
\label{ameanpo}
{\bar a}^\gamma := \frac{1}{T_\gamma}\int_0^{T_\gamma}a\left(\Phi^t_H
(\bi{p},\bi{x})\right)\ \rmd t \ ,\qquad {\rm with}\ (\bi{p},\bi{x})
\in\gamma\ .
\end{equation}
We remark that a heuristic derivation of an analogous identity to (4.4)
was given by Hannay and Ozorio de Almeida \cite{HanOzo84}.

One can apply the relation (\ref{ameanbypo}) to determine the leading
semiclassical behaviour of (\ref{posumest}) once one has chosen a 
suitable observable $a$ whose average ${\bar a}^\gamma$ yields the
quantity $(\tr d_\gamma)^2$ appearing in (\ref{posumest}). This, however,
can only be achieved in an indirect manner. Our choice of the observable
requires to recall the semiclassical time evolution of the spin degrees
of freedom along the trajectories of the classical flow $\Phi_H^t$.
Let $d(\bi{p},\bi{x},t)\in$ SU(2) denote the solution of the spin
transport equation \cite{BolKep98,BolKep99}
\begin{equation}
\label{spintrans}
\dot{d}(\bi{p},\bi{x},t) 
+ \rmi \,M(\Phi_H^t(\bi{p},\bi{x})) \ d(\bi{p},\bi{x},t) = 0 
\ , \quad
d(\bi{p},\bi{x},0) = \mathmybb{1}_2 \ ,
\end{equation}
where the time derivative is understood along the trajectory  
$\Phi_H^{t} (\bi{p},\bi{x})$. $M$ is a certain hermitian and traceless 
$2\times 2$--matrix valued function on $\Omega_E$, whose precise form 
depends on the quantum Hamiltonian under consideration, see \cite{BolKep99} 
for details. Geometrically, $d(\bi{p},\bi{x},t)$ can also be interpreted as 
a parallel transporter in some vector bundle so that 
$d(\bi{p},\bi{x},T_\gamma)$, with $(\bi{p},\bi{x})\in\gamma$, is the 
holonomy associated with the periodic orbit $\gamma$. Since its trace is 
invariant under a shift of the initial point $(\bi{p},\bi{x})\in\gamma$ 
along the orbit, one can introduce the notation $\tr d_\gamma := 
\tr d(\bi{p},\bi{x},T_\gamma)$. We are thus in a position to define the 
observable
\begin{equation}
\label{adef}
a(\bi{p},\bi{x},t) := [\tr d(\bi{p},\bi{x},t)]^2 \ ,
\end{equation}
which is a function on $\Omega_E$ that in addition depends on a parameter 
$t$. Due to the above remark concerning the interpretation of 
$d(\bi{p},\bi{x},T_\gamma)$ as a holonomy, the average (\ref{ameanpo}) of 
this observable along a periodic orbit $\gamma$, when $t=T_\gamma$ is chosen, 
yields 
\begin{equation}
\label{ameanpo1}
{\bar a}^\gamma (T_\gamma) = \frac{1}{T_\gamma}\int_0^{T_\gamma}a
\left(\Phi^{t'}_H(\bi{p},\bi{x}),T_\gamma\right)\ \rmd t' 
= (\tr d_\gamma)^2\ ,
\end{equation}
for any $(\bi{p},\bi{x})\in\gamma$. Without the choice $t=T_\gamma$,
however, ${\bar a}^\gamma (t)$ is not related to $(\tr d_\gamma)^2$.
We can hence now employ (\ref{ameanbypo}) to deduce the asymptotic relation
\begin{equation}
\label{aoftasym}
\sum_{\gamma,\,T_\gamma\leq T}T_\gamma^2\,{\bar a}^\gamma
(t)\,p_\gamma \sim \frac{1}{2}T^2\,{\bar a}^E(t)\ ,\quad T\to\infty\ ,
\end{equation}
which is valid for any $t$. Notice that here we have introduced an extra
power of $T_\gamma$ in the same manner as in (\ref{posumk})--(\ref{asymposum}).
We now differentiate with respect to $T$,
\begin{equation}
\label{diffaoftasym}
\sum_{\gamma}T_\gamma^2\,{\bar a}^\gamma(t)\,p_\gamma\,\delta(T-T_\gamma) 
\sim T\,{\bar a}^E(t)\ ,\quad T\to\infty\ ,
\end{equation}
and then choose $t=T$. Together with (\ref{ameanpo1}) this allows to 
conclude that
\begin{equation}
\label{diffaoftasym1}
\sum_{\gamma}T_\gamma^2\,(\tr d_\gamma)^2\,p_\gamma\,\delta (T-T_\gamma) 
\sim T\,{\bar a}^E(T)\ ,\quad T\to\infty\ .
\end{equation}
Thus, at this point we have obtained the asymptotic relation 
\begin{equation}
\label{KdiagZw}
K_2^{diag}(\tau;T_H) \sim \bar{g} \, \tau \, \bar{a}^E(\tau T_H) 
\end{equation}
for the diagonal form factor (\ref{diagformf}) in the limit $T_H \to \infty$,
$\tau \to 0$ such that $\tau T_H \to \infty$.

The remaining task therefore consists of determining the asymptotics of 
$\bar{a}^E(T)$ as $T\to\infty$. In order to achieve this one has to go back 
to the representation
\begin{equation}
\label{baraE}
\bar{a}^E(T) = \int_{\Omega_E} \left[ \tr d(\bi{p},\bi{x},T) \right]^2 
\ \rmd \mu_{E}(\bi{p},\bi{x}) 
\end{equation}
of $\bar{a}^E(T)$ as an average over phase space. Since the $T$-dependence 
involves $d(\bi{p},\bi{x},T)$ one might anticipate that the limit 
$T\to\infty$ of (\ref{baraE}) depends on certain ergodic properties of the 
spin dynamics. The latter being considered along trajectories of the 
translational dynamics, one hence has to combine both dynamics in a suitable 
way. In ergodic theory the relevant construction is known as a {\it skew 
product}, see, e.g., \cite{CorFomSin82}. Appropriate ergodic properties of 
the skew product dynamics will then allow for a determination of the 
asymptotic behaviour of (\ref{baraE}). Let us therefore now construct
the skew product of translational and spin dynamics. 
To this end one defines a flow $Y^t$ on the product phase space ${\cal M} := 
\Omega_E\times {\rm SU}(2)$ in the following way,
\begin{equation}
  Y^t \left( (\bi{p},\bi{x}), g \right) :=
  \left( \Phi^t_H (\bi{p},\bi{x}), d(\bi{p},\bi{x},t) g \right)
\end{equation}
for $(\bi{p},\bi{x}) \in \Omega_E$ and $g \in {\rm SU}(2)$. The initial
condition $Y^0 ((\bi{p},\bi{x}), g) = ((\bi{p},\bi{x}), g)$ is obviously
fulfilled, and the composition law $Y^{t+t^{\prime}} = Y^t \circ 
Y^{t^{\prime}}$ immediately follows from the relation
\begin{equation}
\label{dcomp}
  d(\bi{p},\bi{x},t+t^{\prime}) 
  = d(\Phi_H^{t^{\prime}}(\bi{p},\bi{x}),t) 
    \ d(\bi{p},\bi{x},t^{\prime}) 
\end{equation}
that can be concluded from (\ref{spintrans}). On ${\cal M}$ one then 
defines the direct product $\mu := \mu_E \times \mu_H$ of Liouville 
measure $\mu_E$ and of the normalized Haar measure $\mu_H$ of ${\rm SU}(2)$. 
We recall that the latter is the {\it unique} normalized left- and 
right-invariant positive Radon measure on the group manifold, see, e.g., 
\cite{BarRac77}. Due to both the invariance of Liouville measure under the 
Hamiltonian flow $\Phi^t_H$ and the left-invariance of Haar measure, the 
product measure $\mu$ is invariant under $Y^t$. A dynamical system of this 
kind is known as an SU(2)-extension of $\Phi^t_H$ or, more generally, a 
skew product. The spin dynamics defined by (\ref{spintrans}) is then called 
a cocycle for $\Phi^t_H$ with values in SU(2). For further information see, 
e.g., \cite{CorFomSin82}.

In addition to the assumptions made for $\Phi^t_H$ in section \ref{3sec}, in 
the following  we will assume that $Y^t$ is (strongly) mixing. This implies
that for any $F \in L^2({\cal M} \times {\cal M})$  
\begin{equation}
\label{skewmix}
\eqalign{
  \lim_{t \to \infty} \int_{\cal M} 
  F \left( Y^t((\bi{p},\bi{x}),g), ((\bi{p},\bi{x}),g) \right) \
  \rmd \mu((\bi{p},\bi{x}),g) \\
  = \int_{\cal M} \int_{\cal M} 
    F \left( ((\bi{p},\bi{x}),g), ((\boldsymbol{\xi},\bi{y}),h) \right) \ 
    \rmd \mu((\bi{p},\bi{x}),g) \, \rmd \mu((\boldsymbol{\xi},\bi{y}),h)  \ .
}
\end{equation}
We remark that usually the mixing property is defined for pairs of functions
$F_1,F_2\in L^2({\cal M})$. However, if one views $L^2({\cal M} \times 
{\cal M})$ as $L^2({\cal M})\otimes L^2({\cal M})$ and introduces a 
tensor-product basis, (\ref{skewmix}) follows immediately because every 
element of this basis fulfills the usual mixing property. If now the 
function $F$ does not depend on the translational degrees of freedom, i.e., 
$F:\,{\rm SU(2)}\times {\rm SU(2)}\to {\mathbb R}$, the mixing property 
(\ref{skewmix}) yields 
\begin{equation}
\fl
\label{mixF}
  \lim_{t \to \infty} \int_{\cal M} 
  F (d(\bi{p},\bi{x},t) g, g) \ \rmd \mu((\bi{p},\bi{x}),g)
  = \int_{\rm SU(2)} \int_{\rm SU(2)} F(g,h) \ \rmd\mu_H(g) \, \rmd\mu_H(h) 
  \ .
\end{equation}
A suitable choice of the function $F$ then allows to determine the asymptotic 
behaviour of $\bar{a}^E(T)$ as $T\to\infty$ from (\ref{mixF}). In order
to achieve this we recall the representation (\ref{baraE}) of $\bar{a}^E(T)$,
which obviously can also be written as 
\begin{equation}
  \bar{a}^E(T) = \int_{\rm SU(2)} \int_{\Omega_E} 
  \left[ \tr ( d(\bi{p},\bi{x},T) g g^{-1}) \right]^2 \   
  \rmd \mu_{E}(\bi{p},\bi{x}) \, \rmd \mu_H(g) \ .
\end{equation}
Upon now defining the function $F(g,h):= [\tr(gh^{-1})]^2$, 
$g,h \in {\rm SU}(2)$, one can employ (\ref{mixF}) to conclude that
\begin{equation}
\eqalign{ 
  \lim_{T \to \infty} \bar{a}^E(T) 
  &= \lim_{T \to \infty} \int_{\cal M} F \left( d(\bi{p},\bi{x},T)g,g \right)
     \ \rmd \mu((\bi{p},\bi{x}),g) \\
  &= \int_{\rm SU(2)} \int_{\rm SU(2)} [\tr(gh^{-1})]^2 
    \ \rmd\mu_H(g) \, \rmd\mu_H(h) \, .
}
\end{equation}
Substituting $g^{\, \prime}=gh^{-1}$ in the inner integral and using the 
right-invariance of $\mu_H$, the integrand does not depend on $h$ any more. 
Thus we obtain 
\begin{equation}
\label{Haarlimit}
  \lim_{T \to \infty} \bar{a}^E(T) 
  = \int_{\rm SU(2)} [\tr g^{\, \prime}]^2\ \rmd\mu_H(g^{\, \prime}) \ ,
\end{equation}
i.e., in the limit $T\to\infty$ the expectation value of 
$[\tr d(\bi{p},\bi{x},T)]^2$, when averaged over phase space, can be 
computed by an average over the group SU(2) with respect to Haar measure. 

The same holds obviously true for any moment of $\tr d(\bi{p},\bi{x},T)$ so 
that the asymptotic distribution of $\tr d(\bi{p},\bi{x},T)$, when the 
initial points $(\bi{p},\bi{x})\in\Omega_E$ are uniformly distributed with 
respect to $\mu_E$, can be computed via
\begin{equation}
\label{trdistr}
\eqalign{ 
\lim_{T\to\infty} \mu_E
  &\left\{ (\bi{p},\bi{x})\in\Omega_E;\ \tr d(\bi{p},\bi{x},T)\in [a,b] 
   \right\}   \\ 
  &=\int_a^b\int_{\rm SU(2)}\delta\left( \tr g - w \right)\ \rmd\mu_H(g)\,
   \rmd w\ .}
\end{equation}
In order to evaluate the integral over SU(2) explicitly we remark that any 
$g \in {\rm SU}(2)$ can be represented as 
\begin{equation}
  g(u) = u_0 \mathmybb{1}_2 + \rmi \boldsymbol{\sigma} \bi{u} \quad 
  {\rm with} \quad u=(u_0,\bi{u})\in{\mathbb R}^4 \quad {\rm and} \quad
  \sum_{j=0}^3 u_j^2 = 1 \ .
\end{equation}
In this parameterization the Haar measure is given by (see, e.g., 
\cite{BarRac77})
\begin{equation}
\label{Haar}
  \rmd\mu_H(g(u)) = \frac{1}{\pi^2} \, 
  \delta \left( \sum_{j=0}^3 u_j^2 - 1 \right)\ \rmd^4 u \ .
\end{equation}
A simple calculation now shows that the distribution (\ref{trdistr}) 
obeys a {\it semicircle law}, i.e., its density reads
\begin{equation}
\label{semicirc}
\fl
p(w) = \int_{\rm SU(2)}\delta\left( \tr g - w \right)\ \rmd\mu_H(g) =
\cases{ \frac{1}{\pi}\,\sqrt{1-\left(\frac{w}{2}\right)^2} & 
 $-2\leq w\leq +2$\ , \\ 0 & else \ .}
\end{equation}
What is required in (\ref{Haarlimit}) is the second moment of the 
distribution (\ref{trdistr}). With the help of (\ref{semicirc}) this
can now easily be computed to yield one, i.e.,
\begin{equation}
\label{baraasym}
\bar a^E(T) \sim 1 \quad {\rm as}\quad T\to\infty\ .
\end{equation}
The integrated version of (\ref{diffaoftasym1}) therefore reads
\begin{equation}
\label{diffaoftasym2}
\sum_{\gamma,\,T_\gamma\leq T}T_\gamma^2\,(\tr d_\gamma)^2\,p_\gamma 
\sim \frac{1}{2}T^2\ ,\quad T\to\infty\ .
\end{equation}
Furthermore, (\ref{KdiagZw}) and (\ref{baraasym}) imply the asymptotic 
behaviour
\begin{equation}
\label{Kdiagresult}
K_2^{diag}(\tau;T_H) \sim \bar{g} \tau 
\end{equation}
of the diagonal form factor in the regime $T_H\to\infty$,
$\tau\to 0$, $\tau T_H\to\infty$.

\section{Summary and Conclusions}
\label{5sec}

In section \ref{3sec} we demonstrated how the two-point form factor for
a quantum system with spin 1/2 can be analyzed semiclassically by making
use of an appropriate trace formula. The diagonal approximation led us
to the periodic-orbit sum (\ref{posumest}) whose asymptotics determines 
the behaviour of the diagonal form factor for small $\tau$ in the
semiclassical limit. Our conclusion (\ref{Kdiagresult}) now has to
be compared with the relevant random matrix results (\ref{ffGSE}) and
(\ref{ffGUE}). To this end we have to distinguish time reversal invariant
quantum Hamiltonians from non-invariant ones. But let us first summarize
our findings. In \ref{appTR} we show that quantum mechanical time reversal 
invariance implies that not only the classical translational dynamics are 
time reversal invariant but, moreover, the spin dynamics behave in such a way 
that also the amplitudes $A_{\gamma,k}$ appearing in the semiclassical trace 
formula are invariant under time reversal. This leads to the 
occurrence of the multiplicities $g_{\gamma,k}$ in the expression 
(\ref{posumest}). Then, when the generic conditions stated at the end of 
section \ref{3sec} are met, we can pull out the factors $g_{\gamma,k}$
from the sum and replace them by either $\bar g=2$, if (quantum mechanical)
time reversal invariance is present, or else by $\bar g=1$. If, furthermore, 
the translational dynamics are ergodic and hyperbolic, the equidistribution 
of periodic orbits allowed us to determine the spin contribution to the 
amplitudes $A_{\gamma,k}$. We further requested the skew product of the 
translational and the spin dynamics to be mixing. This then enabled us to 
identify the distribution of the traces of the spin-transport matrices and 
to calculate its second moment, which enters through the amplitudes 
$A_{\gamma,k}$. 

Hence, if quantum mechanical time reversal invariance is absent, the diagonal
approximation (\ref{Kdiagresult}) states that $K_2^{diag}(\tau;T_H)\sim\tau$, 
which is identical with the diagonal approximation in the case of 
Schr\"odinger operators \cite{Ber85} and coincides with the small-$\tau$ 
asymptotics of the GUE-form factor (\ref{ffGUE}). If, however, the quantum 
Hamiltonian is invariant under time reversal so that we have to choose 
$\bar{g}=2$, the diagonal approximation becomes $K_2^{diag}(\tau;T_H)\sim 
2\tau$. According to (\ref{ffmod}) Kramers' degeneracy then forces us to 
compare the random matrix form factor with the modified semiclassical result
\begin{equation}
  \widetilde{K}_2^{diag}(\tau;T_H)
  = \frac{1}{2} K_2^{diag} \left( \frac{\tau}{2};T_H \right) 
  \sim \frac{1}{2} \tau \ ,
\end{equation}
which now coincides with the small-$\tau$ asymptotics of the GSE-form factor 
(\ref{ffGSE}). In both cases this is exactly the behaviour that is predicted 
by the conjecture of Bohigas, Giannoni and Schmit as stated in 
(\ref{BGSnoT}) and (\ref{BGSwithT}). 

\ack
We would like to thank Prof F Haake for an important remark on Kramers' 
degeneracy. S K would like to thank Prof J P Keating, Dr J Marklof 
and Dr J M Robbins for helpful discussions. S K also acknowledges financial 
support from Deutscher Akademischer Austauschdienst under grant number 
D/99/02553.

\appendix

\section{Classical and quantum mechanical time reversal}
\label{appTR}

In this appendix we are going to discuss the relation between classical 
and quantum mechanical time reversal for systems with spin 1/2. We restrict 
our discussion to the conventional time reversal operation that leaves the 
position coordinates unchanged and reverses momentum and spin coordinates 
as well as the time $t$. The discussion of generalized time reversal 
operators that combine conventional time reversal with another unitary 
symmetry operation is analogous, see \cite{Haa91} for examples. To be 
specific we consider as a quantum Hamiltonian for a particle of mass $m$, 
charge $e$ and spin 1/2 either a Dirac-Hamiltonian
\begin{equation}
\label{HDirac}
\hat H_D  = c\boldsymbol{\alpha} \left( \frac{\hbar}{\rmi}\nabla - 
\frac{e}{c}\bi{A}(\bi{x}) \right) + \beta mc^2 + e\varphi(\bi{x}) \ ,
\end{equation}
or a (generalized) Pauli-Hamiltonian
\begin{equation}
\label{HPauli}
\hat H_P  = \hat H_S \, \mathmybb{1}_2 + \hbar\,\boldsymbol{\sigma}\bi{C} 
\left( \frac{\hbar}{\rmi} \nabla,\bi{x} \right) \ . 
\end{equation}
In the relativistic case (\ref{HDirac}) the Dirac algebra is realized by
the $4\times 4$ matrices  
\begin{equation}
  \boldsymbol{\alpha} = 
  \left( \begin{array}{cc} 0 & \boldsymbol{\sigma} \\
			   \boldsymbol{\sigma} & 0 \end{array} \right)
  \ , \quad \beta = 
  \left( \begin{array}{cc} \mathmybb{1}_2 & 0 \\ 0 & -\mathmybb{1}_2 
  \end{array} \right)
  \ ,
\end{equation}
where $\boldsymbol{\sigma}$ denotes the vector of Pauli matrices and 
$\mathmybb{1}_2$ is a $2\times 2$ unit matrix. The non-relativistic Hamiltonian
(\ref{HPauli}) is composed of a Schr\"odinger operator $\hat H_S$ and a 
coupling term of spin to the translational degrees of freedom. The 
latter has to be understood as the quantization of some 
${\mathbb R}^3$-valued function $\bi{C}(\bi{p},\bi{x})$ on phase space. 
For example, this can be a magnetic field, i.e., $\bi{C}_B(\bi{p},\bi{x}) 
= - \frac{e}{2mc}\,\bi{B}(\bi{x})$, or a spin-orbit coupling term 
$\bi{C}_{so}(\bi{p},\bi{x}) = \frac{1}{4m^2c^2|\bi{x}|}
\frac{\rmd V(|\bi{x}|)}{\rmd |\bi{x}|}\,(\bi{x}\times \bi{p})$. 

For systems with spin 1/2 the operator of time reversal is given by
\begin{equation}
\label{Tdef}
\hat T := \rme^{\rmi\frac{\pi}{2}\sigma_y}\,\hat K = \rmi\,\sigma_y\hat K \ ,
\end{equation}
where $\hat K$ is the operator of complex conjugation in position 
representation, see \cite{Haa91}. In the relativistic case, where $\hat T$
has to act on four component spinors, (\ref{Tdef}) shall mean a block 
diagonal $4\times 4$ matrix with two copies of (\ref{Tdef}) in the diagonal
blocks. A quantum system with Hamiltonian $\hat H$ to be time reversal 
invariant requires $\hat T\hat H\hat T^{-1} = \hat H$. Thus, in the case 
of a Dirac-Hamiltonian (\ref{HDirac})
\begin{equation}
\hat T\hat H_D\hat T^{-1} = c (-\boldsymbol{\alpha}) 
\left( - \frac{\hbar}{\rmi} \nabla - \frac{e}{c} \bi{A}(\bi{x}) \right) 
+ \beta mc^2 + e \varphi(\bi{x}) 
\end{equation}
shows that conventional time reversal invariance is equivalent to 
the absence of magnetic forces. 
For the Pauli-Hamiltonian (\ref{HPauli}) to commute 
with $\hat T$ we first need the Schr\"odinger operator $\hat H_S$ to be 
time reversal invariant, i.e., $\hat K\hat H_S\hat K=\hat H_S$. In 
addition, the condition
\begin{equation}
\hat T \boldsymbol{\sigma}\bi{C}\left( \frac{\hbar}{\rmi}\nabla,\bi{x} 
\right) \hat T^{-1} = -\boldsymbol{\sigma}\bi{C}\left( - \frac{\hbar}{\rmi} 
\nabla,\bi{x} \right) \stackrel{!}{=} \boldsymbol{\sigma}\bi{C}\left(
\frac{\hbar}{\rmi}\nabla,\bi{x} \right) 
\end{equation}
has to be met, i.e., the coupling term $\bi{C}(\bi{p},\bi{x})$ must be an 
odd function of momentum $\bi{p}$. This requirement is fulfilled by the 
spin-orbit coupling term $\bi{C}_{so}$, but is violated by the coupling 
$\bi{C}_B$ to an external magnetic field. In both the relativistic and
the non-relativistic situation, however, even the presence of a magnetic 
field might allow for the existence of an anti-unitary operator representing 
a generalized time reversal symmetry that commutes with the Hamiltonian, 
see \cite{Wig32,Haa91}.

We now want to investigate the implications of a quantum mechanical time 
reversal invariance for the semiclassical analysis of the form factor.
The first obvious consequence is a time reversal invariance of the
classical translational dynamics. But furthermore, also the spin
dynamics, governed by the spin transport equation (\ref{spintrans}),
exhibits a certain kind of symmetry under time reversal. In order to 
discuss the latter, one has to study the behaviour of the matrix $M$
entering (\ref{spintrans}) under $\bi{x}\mapsto\bi{x}$, $\bi{p}\mapsto
-\bi{p}$. For a Pauli-Hamiltonian $M$ is given by $\boldsymbol{\sigma} 
\bi{C}(\bi{p},\bi{x})$, where $\bi{C}(\bi{p},\bi{x})$ is defined as above, 
and for a Dirac-Hamiltonian with no magnetic field one finds 
$M=g(|\bi{p}|)\,\bi{x}\times\bi{p}$, see \cite{BolKep99} for details. In
each case quantum mechanical time reversal invariance therefore implies 
$M(-\bi{p},\bi{x}) = -M(\bi{p},\bi{x})$. Substituting now $t\mapsto -t$ 
and $\bi{p}\mapsto -\bi{p}$ in (\ref{spintrans}) one obtains
\begin{equation}
  - \dot{d}(-\bi{p},\bi{x},-t) 
  + \rmi \,M(\Phi_H^{-t}(-\bi{p},\bi{x})) \ d(-\bi{p},\bi{x},-t) = 0 \ .
\end{equation}
Since the translational dynamics are time reversal invariant, $M$ being
odd in the momentum variable leads to
\begin{equation}
  \dot{d}(-\bi{p},\bi{x},-t) 
  + \rmi \,M(\Phi_H^{t}(\bi{p},\bi{x})) \ d(-\bi{p},\bi{x},-t) = 0 \ ,
\end{equation}
and therefore $d(-\bi{p},\bi{x},-T_{\gamma}) = d(\bi{p},\bi{x},T_{\gamma})$.
Moreover, the fact that $d(\bi{p},\bi{x},T_{\gamma})\in {\rm SU(2)}$ is a 
holonomy implies $d(-\bi{p},\bi{x},-T_{\gamma}) = 
[d(-\bi{p},\bi{x},T_{\gamma})]^{-1}$ so that finally 
$\tr d(-\bi{p},\bi{x},-T_{\gamma}) = \tr d(\bi{p},\bi{x},T_{\gamma})$. 
Altogether the above considerations confirm that the presence of a quantum 
mechanical time reversal invariance implies that a primitive periodic orbit 
$\gamma$ and its time reversed partner share identical amplitudes 
$A_{\gamma,k}$.

\section{Equidistribution of long periodic orbits}
\label{appsumrules}

In this appendix we want to show how the periodic-orbit representation
(\ref{ameanbypo}) of Liouville measure can be obtained from equidistribution
properties of periodic orbits. The basic reference for the following is
\cite{ParPol90}. Our assumptions on the flow $\Phi^t_H$ are as stated 
in \sref{4sec}. 

In order to proceed further we first have to introduce some notation.
Let ${\cal B}$ denote the set of all $\Phi^t_H$-invariant Borel probability
measures on $\Omega_E$. Any $\mu\in {\cal B}$ can be associated a metric
entropy $h_\mu$. Then for any H\"older-continuous observable
$f\in C^\alpha (\Omega_E)$, with some $\alpha >0$, the topological pressure 
is defined as
\begin{equation}
\label{toppress}
P(f) := \sup\left\{ h_\mu + \int_{\Omega_E}f\ \rmd\mu\, 
;\ \mu\in{\cal B} \right\} \ .
\end{equation}
The supremum is attained for a unique measure $\mu_f$, which is called {\it
equilibrium measure} for the observable $f$. The periodic orbits of the flow 
$\Phi^t_H$ are then equidistributed with respect to $\mu_f$ in the following 
sense, 
\begin{equation}
\label{eqdisg}
\int_{\Omega_E}a(\bi{p},\bi{x})\ \rmd\mu_f (\bi{p},\bi{x}) = 
\lim_{T\to\infty} \frac{\sum_{\gamma,\,T_\gamma\leq T}T_\gamma\,
{\bar a}^\gamma\,\exp(T_\gamma{\bar f}^\gamma)}
{\sum_{\gamma,\,T_\gamma\leq T}T_\gamma\,\exp(T_\gamma{\bar f}^\gamma)}\ ,
\end{equation}
for every $a\in C(\Omega_E)$. The averages over periodic orbits are defined 
as in (\ref{ameanpo}).

In a next step one has to identify Liouville measure as the equilibrium
measure of some observable $f$. To this end we consider the tangential
map $D\Phi^t_H$ restricted to the unstable subbundle $E^u$ of the
tangent bundle $T\Omega_E$ and define
\begin{equation}
\label{defg}
f(\bi{p},\bi{x}) := -\frac{\rmd}{\rmd t}\log\det \left. D\Phi^t_H
(\bi{p},\bi{x})\right|_{E^u, t=0} \ . 
\end{equation}
A direct calculation then yields
\begin{equation}
\label{gmeanpo}
{\bar f}^\gamma = -\frac{1}{T_\gamma}\sum_{j=1}^{d-1}u_{\gamma,j}
\quad {\rm so\ that}\quad \rme^{T_\gamma{\bar f}^\gamma} = p_\gamma \ ,
\end{equation}
compare (\ref{DetMasy}). Furthermore, the equilibrium measure associated
with the observable (\ref{defg}) is called {\it Sinai-Ruelle-Bowen} 
measure $\mu_{SRB}$, for which it is known that for every $a\in C(\Omega_E)$
\begin{equation}
\label{SRBprop}
\lim_{T\to\infty}\frac{1}{T}\int_0^T a\left(\Phi^t_H(\bi{p},\bi{x})\right)
\ \rmd t = \int_{\Omega_E} a(\bi{p}',\bi{x}')\ \rmd\mu_{SRB}(\bi{p}',\bi{x}')
\end{equation}
holds for a set of initial conditions $(\bi{p},\bi{x})\in\Omega_E$ with 
positive Liouville measure, see \cite{BowRue75} for more information.
Since $\Phi^t_H$ is supposed to be ergodic with respect to $\mu_E$, one
concludes that $\mu_{SRB}=\mu_E$.

What is still lacking is an asymptotic estimate of the denominator on
the right-hand side of (\ref{eqdisg}). In order to obtain this we first
have to introduce a regularization in that we multiply the observable
(\ref{defg}) by some factor $\beta<1$. Then we appeal to the relation
\begin{equation}
\label{asymreg}
\sum_{\gamma\atop T-\varepsilon\leq T_\gamma\leq T+\varepsilon}T_\gamma
\,p_\gamma^\beta \sim \frac{\rme^{P(\beta f)T}}{P(\beta f)}\,\left[
\rme^{P(\beta f)2\varepsilon} - 1 \right]\ ,\quad T\to\infty\ ,
\end{equation}
see \cite{ParPol90}. Since $P(f)=0$, in the limit $\beta\to 1$ one obtains
for $k\in {\mathbb N}$
\begin{equation}
\label{posumk}
\sum_{\gamma\atop T-\varepsilon\leq T_\gamma\leq T+\varepsilon}T_\gamma^k
\,p_\gamma \sim 2\varepsilon\,T^{k-1}\ ,\quad T\to\infty\ .
\end{equation}
Replacing now $\varepsilon$ by $T$, followed by the rescaling 
$2T\mapsto T$, this implies
\begin{equation}
\label{asymposum}
\sum_{\gamma\atop T_\gamma\leq T}T_\gamma^k\,p_\gamma \sim 
2^{1-k}\,T^k\ ,\quad T\to\infty\ .
\end{equation}
Introducing then the relations (\ref{gmeanpo}) and (\ref{asymposum})
in (\ref{eqdisg}) finally yields the periodic-orbit representation
(\ref{ameanbypo}) of Liouville measure.

\end{document}